\documentclass[11pt,a4paper]{article}

\usepackage{exscale,german,amsthm}
\usepackage[intlimits]{amsmath}
\usepackage[english]{babel}
\usepackage[cp850]{inputenc}
\usepackage{epsfig}
\usepackage{enumerate}

\newcommand{\bol}[1]{\mbox{\boldmath$#1$}}

\newcommand{\bSigma}{\bol{\Sigma}}

\newcommand{\bmu}{\bol{\mu}}
\newcommand{\tbm}{\tilde{\bol{\mu}}}

\newcommand{\bx}{\mathbf{X}}
\newcommand{\bQ}{\mathbf{Q}}

\newcommand{\bP}{\mathbf{P}}

\newcommand{\bw}{\mathbf{w}}

\newcommand{\bii}{\mathbf{1}}

\newcommand{\bA}{\bol{A}}

\newcommand\Tstrut{\rule{0pt}{5ex}}
\newcommand\Bstrut{\rule[-4ex]{0pt}{0pt}}

\newtheorem{theorem}{Theorem}

\newtheorem{proposition}{Proposition}

\newfont{\tabfont}{cmr7 at 7pt}

\begin{document}

\begin{center}
\vspace*{2cm} \noindent {\bf \large On the Equivalence of Quadratic Optimization Problems Commonly Used in Portfolio Theory}\\
\vspace{1cm} \noindent {\sc  Taras Bodnar$^{a}$, Nestor Parolya$^{b}$ and Wolfgang Schmid$^{b,*}$
\footnote{$^*$ Corresponding author. E-mail address: schmid@euv-frankfurt-o.de} }\\

\vspace{1cm} {\it \footnotesize  $^a$
Department of Mathematics, Humboldt-University of Berlin, Unter den Linden 6, 10099 Berlin, Germany} \\
{\it \footnotesize  $^b$
Department of Statistics, European University Viadrina, PO Box 1786, 15207 Frankfurt (Oder), Germany} \\
\end{center}

\begin{abstract}
In the paper, we consider three quadratic optimization problems which are frequently applied in portfolio theory, i.e, the Markowitz mean-variance problem
as well as the problems based on the mean-variance utility function and the quadratic utility. Conditions are derived under which the solutions of these three optimization procedures coincide and are lying on the efficient frontier, the set of mean-variance optimal portfolios. It is shown that the solutions of the Markowitz optimization problem and the quadratic utility problem are not always mean-variance efficient.

The conditions for the mean-variance efficiency of the solutions depend on the unknown parameters of the asset returns. We deal with the problem of parameter uncertainty in detail and derive the probabilities that the estimated solutions of the Markowitz problem and the quadratic utility problem are mean-variance efficient.
Because these probabilities deviate from one the above mentioned quadratic optimization problems are not stochastically equivalent. The obtained results are illustrated by an empirical study.

\end{abstract}

\vspace{0.7cm}

\newpage

\section{Introduction}

\noindent The portfolio selection problem is one of the most interesting and important topics in investment theory which plays an important role in financial research nowadays. A huge number of papers are dealing with a mean-variance portfolio and its characteristics (e.g., Markowitz (1952), Tobin (1958), Merton (1972), Gibbons et al. (1989), Jobson and Korkie (1989), White (1998), Korkie and Turtle (2002), Okhrin and Schmid (2006), Basak et al. (2009)). Currently, the mean-variance analysis derived by Markowitz (1952) is of great importance for both researchers and practitioners on the financial sector (cf. Litterman (2003), Brandt (2010)). It is the first systematic treatment of an investor's conflicting objectives, a high return versus a low risk. The mean-variance analysis is based on a quadratic parametric optimization model for the single-period portfolio selection problem. An explicit analytical solution is obtained for an investor trying to maximize his expected wealth without exceeding a predetermined risk level and an investor trying to minimize his risk ensuring a predetermined wealth respectively. Alternatively, two other quadratic optimization procedures have been recently considered in literature (see, e.g., Ingersoll (1987), Brandt and Santa-Clara (2006), Bodnar and Schmid (2008b, 2009), \c{C}elikyurt and \"{O}zekici (2007), Fu et al., (2010), Cesarone et al.,(2011)). Note that a quadratic optimization problem satisfies the Bernoulli principle without imposing the assumption of normality on the distribution of the asset returns (see Tobin (1958)).

In the following we consider an investor who holds a portfolio consisting of $k$ assets. The $k$-dimensional vector of the asset returns is denoted by $\bx=(X_1,\ldots,X_k)^\prime$. Let $\bmu$ be the mean vector of $\bx$, i.e. $E(\bx)=\bmu$, and $\bSigma=Cov(\bx)$ be its covariance matrix. We assume that $\bSigma$ is positive definite.  The vector $\bw=(w_1,\ldots,w_k)^\prime$ and the vector $\bii=(1,\ldots,1)^\prime$ denote the $k$-dimensional vector of portfolio weights and the $k$-dimensional vector of ones, respectively.

The first problem considered in the paper is the so-called mean-variance optimization problem suggested in the seminal paper of Markowitz (1952). Kroll et al. (1984) showed that the mean-variance optimal portfolio also possesses a maximum expected utility or is very close to this value for utility functions commonly used in financial literature. In the optimization problem of Markowitz an investor is considered who minimizes the portfolio risk for a given level of the expected return. It is formulated as
\begin{equation}\label{problem1}
\bw^\prime\bSigma\bw\rightarrow\min~~~\text{subject to}~~~\bw^\prime\bmu=\mu_0, ~~~\bw^\prime\bii=1.
\end{equation}
Throughout the paper we refer to (\ref{problem1}) as the Markowitz mean-variance (M) optimization problem.

Secondly, we consider an investor who maximizes the mean-variance utility function. The corresponding optimization procedure, to which we refer as the mean-variance utility (MVU) problem, is given by
\begin{equation}\label{problem2}
\bw^\prime\bmu-\frac{\alpha}{2}\bw^\prime\bSigma\bw\rightarrow\max~~~\text{subject to}~~~\bw^\prime\bii=1.
\end{equation}
The quantity $\alpha>0$ is the slope parameter of the quadratic utility function. In general it is not equal to the risk aversion coefficient as defined by Pratt (1964). Moreover, the quantity $\alpha$ determines the risk of the investor under the assumption of a normal distribution. Note that the mean-variance utility function assuming normally distributed asset returns coincides with the exponential (or CARA) utility function $U(W)=-e^{-\alpha W}/{\alpha}$ where $W$ denotes the investor's wealth (see, e.g., Merton (1969), \c{C}elikyurt and \"{O}zekici (2009)). The comparison of different utility functions can be found in Yu et al. (2009).

The portfolio that maximizes (\ref{problem2}) is called the expected utility (EU) portfolio. It has been heavily discussed  in financial literature recently (see, e.g., Ingersoll (1987), Okhrin and Schmid (2006), Bodnar and Schmid (2008b, 2009)). While Okhrin and Schmid (2006) obtained the first two moments of the estimated weights of the EU portfolio, Bodnar and Schmid (2008b, 2009) derived the sample distributions of its estimated expected return and variance.


Finally, our third optimization procedure is, usually, considered in the multi-stage portfolio selection theory (see, e.g., Brandt and Santa-Clara (2006)). It is based on the following quadratic utility function $U(W)=W-\dfrac{\tilde\alpha}{2}W^2$, where $W$ denotes the investor's wealth and $\tilde \alpha>0$ determines the investor attitude towards risk. It is formulated as
\begin{equation}\label{problem3}
E(U(W))=E(W)-\frac{\tilde\alpha}{2}E(W^2)=W_0\bw^\prime\tbm-\frac{\tilde{\alpha}}{2}W_0^2\bw^\prime\bA\bw\rightarrow\max~~~\text{subject to}~~~\bw^\prime\bii=1,
\end{equation}
where $\bA=\bSigma+\tbm\tbm^\prime$ with $\tbm=\bii+\bmu$ and $W=W_0(1+\bw^\prime \mathbf{X})$ where $W_0$ is the initial value of the wealth. Without loss of generality we put $W_0=1$. We refer to (\ref{problem3}) as the quadratic utility (QU) optimization problem.

In Table 1 we summarize the optimization problems (\ref{problem1})-(\ref{problem3}) and present their solutions. Although these three quadratic optimization problems, i.e, the Markowitz mean-variance problem (M) as well as the problems based on the mean-variance utility function (MVU) and the quadratic utility (QU), are intensively discussed in financial theory and financial practice we have not found papers which compare the three considered quadratic optimization problems with each other. The comparison of different existing models is always meaningful to understand
the theory more deeply. The only papers, where a similar problem has been briefly discussed, are the papers written by Li and Ng (2000) and Leippold et al. (2004). In these papers it was pointed out that the three quadratic optimization problems are mathematically equivalent but not economically. Although their solutions are lying on the same set in the mean-variance space, we show that they are not obviously the same.$^1$\footnote{$^1$The solutions of the three considered optimization problems lie on a parabola in the mean-variance space. However, only using the (MVU) optimization problem, the investor always gets a mean-variance efficient portfolio because only in this case all solutions lie on the upper part of the parabola. This is not the case for the optimization problem (M) and (QU) since in these cases we can get portfolios from the lower part of the parabola.
Nevertheless, they can be made equivalent by adding a corresponding constraint in Markowitz's problem as well as in the problem based on the quadratic utility function. However, the constraints themselves depend on the unknown quantities and, hence, they cannot be checked in practice. As a result, the three problems are mathematically equivalent but not from a stochastic point of view.}

\vspace{0.1cm}
\begin{center}
\begin{table}[h]\small
\begin{tabular}{|l|l|}
\hline
Optimization Problem & Solution\\
\hline
$(1)~~\bw^\prime\bSigma\bw\rightarrow\min~~~\text{subject to}~~~\bw^\prime\bmu=\mu_0, ~~~\bw^\prime\bii=1$&$\bw^*=\dfrac{a-\mu_0b}{ac-b^2}\bSigma^{-1}\bii+\dfrac{(\mu_0c-b)}{ac-b^2}\bSigma^{-1}\bmu$\,\Tstrut\Bstrut\\
\hline
$(2)~~\bw^\prime\bmu-\dfrac{\alpha}{2}\bw^\prime\bSigma\bw\rightarrow\max~~~\text{subject to}~~~\bw^\prime\bii=1$&$\bw^*=\dfrac{\bSigma^{-1}\bii}{\bii^\prime\bSigma^{-1}\bii}+\alpha\left(\bSigma^{-1}-\displaystyle\frac{\bSigma^{-1}\bii\bii^\prime\bSigma^{-1}}{\bii^\prime\bSigma^{-1}\bii}
\right)\bmu\,$\Tstrut\Bstrut\\
\hline
$(3)~~\bw^\prime\tbm-\dfrac{\tilde{\alpha}}{2}\bw^\prime\bA\bw\rightarrow\max~~~\text{subject to}~~~\bw^\prime\bii=1$&$\bw^*=\displaystyle\frac{\bA^{-1}\bii}{\bii^\prime\bA^{-1}\bii}+\tilde{\alpha}^{-1}\left(\bA^{-1}-\displaystyle\frac{\bA^{-1}\bii\bii^\prime\bA^{-1}}{\bii^\prime\bA^{-1}\bii}
\right)\tbm\,$\Tstrut\Bstrut\\
\hline
\end{tabular}
\caption{Solutions of the optimization problems (M), (MVU) and (QU). The constants are defined by $a=\bmu\bSigma^{-1}\bmu^\prime$, $b=\bii^\prime\bSigma^{-1}\bmu$, and $c=\bii^\prime\bSigma^{-1}\bii$.}
\end{table}
\end{center}

In this paper these results are extended in several directions. First, we derive sufficient conditions for the mathematical equivalence which are new and have not been previously discussed in financial literature to the best of our knowledge. These conditions only depend on the parameters of the asset returns. The investor only has to change his risk
aversion parameter to obtain an equivalent solution for another optimization problem. Second, the influence of parameter uncertainty on the three optimization problems is investigated. It holds that
the parameters of the asset returns, namely the mean vector and the covariance matrix, are usually unknown in practice. As a result, the sufficient conditions for the equivalence cannot be directly verified in practice since they are functions of these two parameters. The question arises whether the solutions of Markowitz's problem and of the quadratic utility problem are always mean-variance efficient. If not then we would like to know how likely are inefficient portfolios obtained. In order to take the parameter uncertainty into account we replace the parameters in the equivalence
conditions by their sample estimators. We derive the distributions of these estimated quantities assuming that the asset returns are independently and identically normally distributed. Although the assumption of normality is heavily criticized in financial literature, it can be applied in the case of the mean-variance investor (see Tu and Zhou (2004)). Furthermore, by the detailed analysis of the probability that the solutions of the optimization problems coincide, we derive a test whether a given solution is mean-variance efficient. Contrary to the classical testing theory for the mean-variance efficiency of a portfolio (see e.g. Gibbons et al. (1989), Britten-Jones (1999), Bodnar and Schmid (2009)), the suggested test on efficiency is constructed under the alternative hypothesis and, consequently, it can be accepted. In order to obtain the finite sample distributions of the estimators and exact tests on the corresponding population quantities, the results of Bodnar and Schmid (2008b, 2009) are used. Moreover, using real data we show that if these parameters are estimated and replaced by the corresponding estimators then the probability of getting an inefficient portfolio by using Markowitz's optimization problem and the optimization problem based on the quadratic utility function can reach $50\%$.


The rest of the paper is organized as follows. In Section 2 we discuss the case with known parameters. In Theorem 1 conditions are derived under which the solution of (\ref{problem1}) coincides with the solution of (\ref{problem2}) and as the solution of (\ref{problem3}) is the same as the solution of (\ref{problem2}), respectively. It appears that these conditions depend on the unknown parameters $\bmu$ and $\bSigma$. In Section 3 the problem of parameter uncertainty is taken into account. We replace the parameters $\bmu$ and $\bSigma$ in (\ref{problem1})-(\ref{problem3}) by their sample estimators. By doing this we get estimators of the corresponding optimal portfolio weights.
 In Theorem 2, we derive the probability that the estimated portfolio based on (\ref{problem1}) (as well as the estimated solution based on (\ref{problem3})) are not mean-variance efficient.
 In Section 3.2 a test theory for the mean-variance efficiency is developed.
An empirical illustration is provided in Section 4. Section 5 presents final remarks. The proof of Theorem 1 is given in the appendix (Section 6).

\section{Portfolio Selection with Known Parameters}
\noindent Throughout this section we assume that both parameters of the asset return distribution, i.e. $\bmu$ and $\bSigma$, are known quantities. Under this assumption Merton (1972) showed that all solutions of (\ref{problem1}) are lying on a parabola in the mean-variance space. The upper part of this parabola is called the efficient frontier, which is the set of all optimal portfolios. The parabola is fully defined by the three parameters $\{a=\bmu\bSigma^{-1}\bmu^\prime,b=\bii^\prime\bSigma^{-1}\bmu,c=\bii^\prime\bSigma^{-1}\bii\}$, which are known as the efficient set constants (see, e.g. Pennacchi (2008)).

Using (\ref{problem2}) Bodnar and Schmid (2008b) suggested an equivalent representation of the efficient frontier given by
\begin{equation}\label{ef}
(R-R_{GMV})^2=\bmu^\prime\bQ\bmu(V-V_{GMV})~~~~\text{with}~~~ \bQ=\bSigma^{-1}-\frac{\bSigma^{-1}\bii\bii^\prime\bSigma^{-1}}{\bii^\prime\bSigma^{-1}\bii}\,,
\end{equation}
where
\begin{equation}\label{rv}
R_{GMV}=\frac{\bii^\prime\bSigma^{-1}\bmu}{\bii^\prime\bSigma^{-1}\bii},~~~~V_{GMV}=\frac{1}{\bii^\prime\bSigma^{-1}\bii}~~~~\text{and}~~~~s=\bmu^\prime\bQ\bmu
\end{equation}
are the expected return and the variance of the global minimum variance portfolio (GMV) which is the portfolio with the smallest variance. The quantity $s=\bmu^\prime\bQ\bmu$ denotes the slope coefficient of the parabola. The set of characteristics $\{R_{GMV},V_{GMV},s\}$ is easier tractable than the set of the constants $\{a,b,c\}$ introduced by Merton (1972). Moreover, Bodnar and Schmid (2008b) showed that the solution of (\ref{problem2}) for $\alpha>0$ coincides with the efficient frontier, i.e. all optimal portfolios obtained by maximizing $\bw^\prime\bmu-\frac{\alpha}{2}\bw^\prime\bSigma\bw$ given $\bw^\prime\bii=1$ are mean-variance efficient.

It is important to note that in general the solution of (\ref{problem1}) as well as the solution of (\ref{problem3}) do not coincide with the efficient frontier. They both lie on the parabola in the mean-variance space, but these sets can be larger or smaller than the efficient frontier. The aim of this section is to provide a comprehensive analysis of this point.

First, we present a short illustration that fully motivates the topic. For this reason, we make use of
\begin{equation}\label{param1}
R_{GMV}=0.014, \quad V_{GMV}=0.0011, \quad \text{and} \;\; s=0.25 \,.
\end{equation}
which are considered as the true parameters of the efficient frontier in this example. In Figure 1 we plot the solutions of the M problem and of the QU problem for $\mu_0>0$ and $\tilde{\alpha}>0$. We observe that the graphs of the solutions of (\ref{problem1}) as well as (\ref{problem3}) include that of (\ref{problem2}). While the graph of the solutions of (\ref{problem2}) is equal to the upper part of the parabola, the efficient frontier, the solutions of (\ref{problem1}) and (\ref{problem3}) may lie on the lower part, respectively. It has to be noted that the lower part of the parabola does not consist of optimal portfolios any longer, since for the same value of the variance we are able to specify a portfolio from the upper part of the parabola which has a larger expected return. Moreover, the lower parts of the corresponding solutions are quite large in both cases, especially in the case of the QU optimization procedure.

In Theorem 1 we state the conditions under which the optimization problems (\ref{problem1}), (\ref{problem2}), and (\ref{problem3}) are equivalent, i.e. they all generate a solution that lies on the efficient frontier.

\begin{theorem}
Let $\bx$ be a random return vector with mean vector $\bmu$ and positive definite covariance matrix $\bSigma$.
\begin{enumerate}[i)]
\item Let $\mu_0 \ge R_{GMV}$ be given. Then a solution of (\ref{problem1}) is a solution of (\ref{problem2}) and it is obtained by choosing $\alpha$ equal to
\begin{equation}\label{a1}
\alpha_{1}=\left\{
  \begin{array}{l l}
   \frac{s}{\mu_0-R_{GMV}} &\quad\text{for}~~~~\mu_0 \ge R_{GMV}\\
   \infty & \quad \text{for}~~~ \mu_0=R_{GMV}
  \end{array} \right.\,.
\end{equation}
If, however, $\mu_0 < R_{GMV}$ then the solution of (\ref{problem1}) does not lie on the efficient frontier and thus it is not a solution of (\ref{problem2}).
\item Let $\tilde{\alpha}^{-1} \ge 1+R_{GMV}$ be given. Then a solution of (\ref{problem3}) is a solution of (\ref{problem2}) and it is obtained by choosing $\alpha$ equal to
\begin{equation}\label{a2}
\alpha_{3}=\left\{
  \begin{array}{l l}
   \frac{1+s}{\tilde{\alpha}^{-1}-1-R_{GMV}} &\quad\text{for}~~~~\tilde{\alpha}^{-1} > 1+R_{GMV}\\
   \infty & \quad \text{for}~~~ \tilde{\alpha}^{-1}=1+R_{GMV}
  \end{array} \right.\,.
\end{equation}
If, however, $\tilde{\alpha}^{-1}<1+R_{GMV}$ then the solution of (\ref{problem3}) does not lie on the efficient frontier and thus it is not a solution of (\ref{problem2}).
\end{enumerate}
\end{theorem}

The proof of Theorem 1 is given in the appendix. For $\alpha_1=\infty$ ($\alpha_3=\infty$) the solution is the global minimum variance portfolio. Similar findings are also given in Leippold et al. (2004) who pointed out that the optimization problems are mathematically equivalent but not economically. However, no condition of the equivalence of (\ref{problem1}), (\ref{problem2}), and (\ref{problem3}) is presented in that paper. If $R_{GMV}$ is a known quantity then the conditions of mean-variance efficiency in Theorem 1a can be directly checked. However, the quantities $\bmu$ and $\bSigma$ are unknown in practice and have to be replaced by the suitable estimators. In this case we are unable to guarantee that the solution of (\ref{problem1}) or/and (\ref{problem3}) lies always on the estimated efficient frontier. The problem is much more difficult in this case and it will be treated in Section 3.

 In the empirical illustration considered in Figure 1, it holds that $\alpha^{-1}_{1}<0$ for $\mu_0\in[0,0.014)$ and $\alpha^{-1}_{3}<0$ for $\tilde{\alpha}\in[0,1.014)$. For these values of $\mu_0$ and $\tilde{\alpha}$ the corresponding solutions of (\ref{problem1}) and (\ref{problem3}) are no longer mean-variance optimal portfolios.

\begin{figure}[ptbh]
\begin{tabular}{p{8.5cm} p{8.5cm}}
$\text{a) \textit{Solution of the M problem}} \hspace{2.6cm}$
\includegraphics[width=7.9cm]{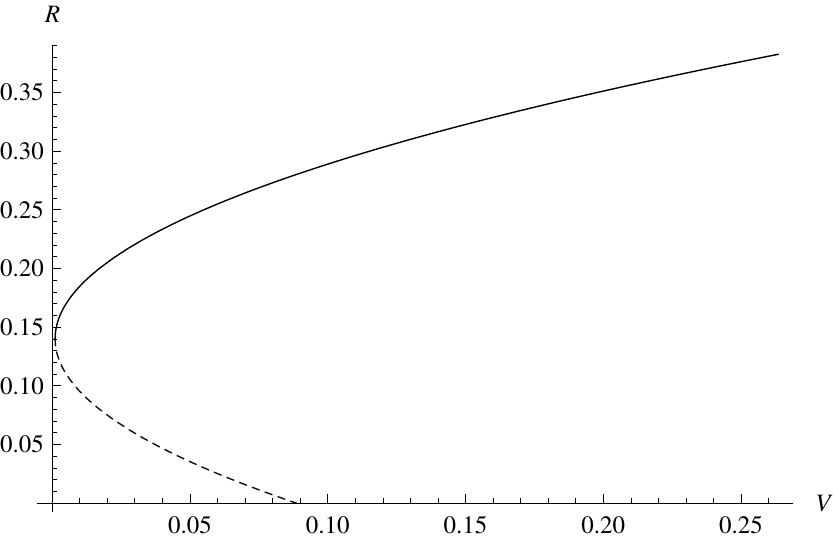}
& $\text{b) \textit{Solution of the QU problem}} \hspace{2.6cm}$ \includegraphics[width=7.9cm]{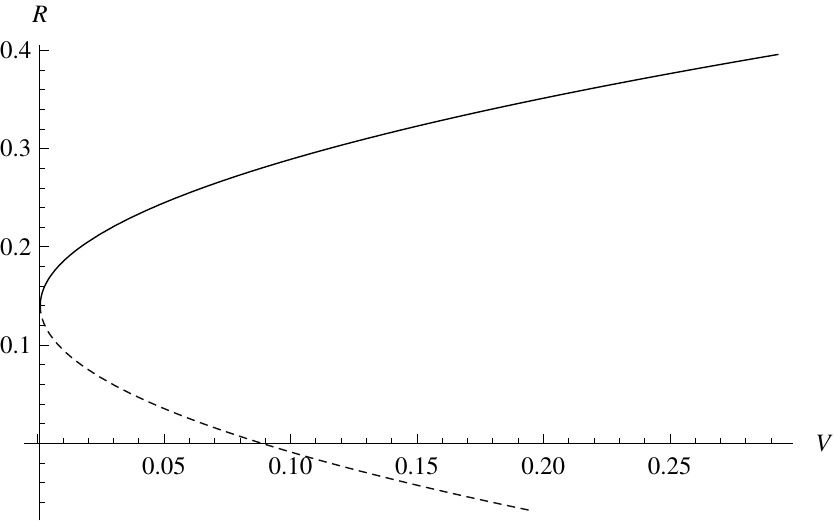}\\
\end{tabular}
\caption{\textit{The solution of the M problem (part a)) and the solution of the QU problem (part b)) in the case of $R_{GMV}=0.014$, $V_{GMV}=0.0011$ and $s=0.25$.
The upper parts of the parabolas are the solution of the MVU optimization problem in both cases.}}%
\label{Fig:his}%
\end{figure}

\section{Influence of Parameter Uncertainty on Portfolio Selection}
\noindent The parameters  of the asset returns, namely $\bmu$ and $\bSigma$, are, usually, unknown in practice. As a result the quantities $R_{GMV}$, $V_{GMV}$, and $s$ cannot be explicitly calculated and the efficient frontier cannot be constructed. The common solution applied in practice is to estimate $\bmu$ and $\bSigma$ by the sample counterparts, i.e.
\begin{equation}\label{est}
\hat{\bmu}=\frac{1}{n}\sum\limits_{j=1}^n\bx_j~~~~\text{and}~~~~\hat{\bSigma}=\frac{1}{n-1}\sum\limits_{j=1}^n(\bx_j-\hat{\bmu})(\bx_j-\hat{\bmu})^\prime\,,
\end{equation}
and to replace $\bmu$ and $\bSigma$ by $\hat{\bmu}$ and $\hat{\bSigma}$ in the corresponding expressions. The estimators of the three parameters of the efficient frontier are given by
\begin{equation}\label{estf}
\hat{R}_{GMV}=\frac{\bii^\prime\hat{\bSigma}^{-1}\hat{\bmu}}{\bii^\prime\hat{\bSigma}^{-1}\bii},~\hat{V}_{GMV}=\frac{1}{\bii^\prime\hat{\bSigma}^{-1}\bii},~\hat{s}=\hat{\bmu}^\prime\hat{\bQ}\hat{\bmu}~~
\text{with}~~\hat{\bQ}=\hat{\bSigma}^{-1}-\frac{\hat{\bSigma}^{-1}\bii\bii^\prime\hat{\bSigma}^{-1}}{\bii^\prime\hat{\bSigma}^{-1}\bii}\,.
\end{equation}
Because $\hat{\bmu}$ and $\hat{\bSigma}$ are random quantities, the estimated characteristics in (\ref{estf}) and the estimator of the efficient frontier in (\ref{ef}) are random as well. Bodnar and Schmid (2008b, 2009) derived the exact distributions of $\hat{R}_{GMV}$, $\hat{V}_{GMV}$, and $\hat{s}$ assuming that the asset returns are independently and identically multivariate normally distributed. Although the assumption of independence and normality are heavily criticized in financial literature, they can be imposed in important practical situations. For instance, Fama (1976) argued that monthly data can be well fitted by the normal distribution, while Tu and Zhou (2004) showed that the effect of heavy tails is relatively small for the mean-variance investor.

Let $\phi(\cdot)$ denote the density function of the standard normal distribution. By $f_{\chi^2_n}(\cdot)$ we denote the density of the $\chi^2$-distribution with $n$ degrees of freedom, while $f_{F_{n_1,n_2,\lambda}}(\cdot)$ stands for the density of the non-central $F$-distribution with $n_1$ and $n_2$ degrees of freedom and the non-centrality parameter $\lambda$. The symbol $f_{N(\mu,\sigma^2)}(.)$ is used for the density function of the normal distribution with mean $\mu$ and variance $\sigma^2$.

Next, for presentation purposes, we summarize some results of Bodnar and Schmid (2008b, 2009) in Proposition 1 where the distributional properties of the three estimated parameters of the efficient frontier are discussed in detail. Namely, the exact joint and marginal distributions of $\hat{R}_{GMV}$, $\hat{V}_{GMV}$, and $\hat{s}$ were derived and they were used for obtaining a confidence region for the efficient frontier. Later on we apply the distributional results of Proposition 1 for other purposes. Using the exact distribution of $\hat{R}_{GMV}$, $\hat{V}_{GMV}$, and $\hat{s}$ we derive the distributions of the test statistics for testing the null hypotheses that the obtained solutions of (\ref{problem1}) or/and (\ref{problem3}) for a given value of $\mu_0$ and $\tilde{\alpha}$, respectively, are mean-variance efficient, i.e. if the following two inequalities $\alpha_1^{-1} \le 0$ and $\alpha_3^{-1} \le 0$ hold under the null hypothesis. Moreover, for given values of $\mu_0$ and $\tilde{\alpha}$ we calculate the probabilities that the estimators of portfolios based on (\ref{problem1}) and (\ref{problem3}) respectively are not mean-variance efficient as functions of $\mu_0$ and $\tilde{\alpha}$ respectively.

\begin{proposition}
Let $\bx$ be a random return vector with mean $\bmu$ and positive definite matrix $\bSigma$ and let $\bx_1,\ldots,\bx_n$ be a random sample independent vectors such that $\bx_i\sim\mathcal{N}_k(\bmu,\bSigma)$ for $i=1,\ldots,n.$ and $n > k$. Then it holds that
\begin{itemize}
\item[a)] $\hat{V}_{GMV}$ is independent of $(\hat{R}_{GMV},\hat{s})$.
\item[ b)] $(n-1){\hat{V}_{GMV}}/{V_{GMV}} \sim \chi^2_{n-k}$.
\item[c)] $\frac{n(n-k+1)}{(n-1)(k-1)} \hat{s}\sim F_{k-1,n-k+1,n\,s}$.
\item[d)] $\hat{R}_{GMV}| \hat{s}=y \sim \mathcal{N}\left(R_{GMV},\frac{1+\frac{n}{n-1}y}{n}V_{GMV}\right)$.
\item[e)] The joint density function of $\hat{R}_{GMV}$, $\hat{V}_{GMV}$, and $\hat{s}$ is given by
\begin{eqnarray*}
&&f_{\hat{R}_{GMV},\hat{V}_{GMV},\hat{s}}(x,y,z)
=\frac{n(n-k+1)}{(k-1)V_{GMV}}
f_{\chi^2_{n-k}}(\frac{n-1}{V_{GMV}}z)\\
&\times&f_{N(R_{GMV},\frac{1+\frac{n}{n-1}y}{n}V_{GMV})}(x)
f_{F_{k-1,n-k+1,n\,s}}(\frac{n(n-k+1)}{(n-1)(k-1)}y) \,.
\end{eqnarray*}
\end{itemize}
\end{proposition}

Recently, a number of papers deals with the application of shrinkage estimators in portfolio analysis (see, e.g., Wang (2005), Okhrin and Schmid (2007, 2008), Basak et al. (2009), Frahm and Memmel (2010)). Under some regularity conditions, these estimators are asymptotically normally distributed. As a result, similar asymptotic results can be obtained to those given in Sections 3.1 and 3.2, where small sample distributions are derived by applying (\ref{est}) for estimating $\bmu$ and $\bSigma$ and the results of Proposition 1.

\subsection{Parameter estimation}
\noindent First, we note that it is impossible in practice to guarantee that the solutions of (\ref{problem1}) and (\ref{problem3}) lie on the efficient frontier, i.e. they coincide with the solution of (\ref{problem2}). In Theorem 1 the corresponding conditions are presented under which the solutions of (\ref{problem1}) and (\ref{problem3}) are lying on the efficient frontier, but both conditions depend on the unknown parameters $R_{GMV}$ and $s$ and, as a result, they cannot be checked in practice. For that reason we are interested in the question whether for a fixed value of $\mu_0$ and for a fixed value of $\tilde{\alpha}$ the estimated solutions of (\ref{problem1}) and (\ref{problem3}) are lying on the estimated efficient frontier, i.e. whether they are mean-variance efficient or not.

Let
\begin{equation}\label{a123es}
\hat{\alpha}^{-1}_{1}=\frac{\mu_0-\hat{R}_{GMV}}{\hat{s}}~~~~\text{and}~~~~\hat{\alpha}^{-1}_{3}=\frac{\tilde{\alpha}^{-1}-1-\hat{R}_{GMV}}{1+\hat{s}}.
\end{equation}
be the estimates of $\alpha^{-1}_{1}$ and $\alpha^{-1}_{3}$ given in (\ref{a1}) and (\ref{a2}), where $\hat{s}$ and $\hat{R}_{GMV}$ are provided in (\ref{estf}). Because $\hat{\alpha}^{-1}_{1}$ and $\hat{\alpha}^{-1}_{3}$ are random variables we consider the probability that $\bP(\hat{\alpha}^{-1}_{1}<0)$ and $\bP(\hat{\alpha}^{-1}_{3}<0)$. These probabilities are derived by using the facts that $\hat{R}_{GMV}| \hat{s}=y \sim \mathcal{N}\left(R_{GMV},\frac{1+\frac{n}{n-1}y}{n}V_{GMV}\right)$ and  $\frac{n(n-k+1)}{(n-1)(k-1)} \hat{s}\sim F_{k-1,n-k+1,n\,s}$ (see Proposition 1). Our results are summarized in Theorem 2, where we use the symbol $\Phi(\cdot)$ for the standard normal distribution function.

\begin{theorem}
Let $\bx$ be a random return vector with mean $\bmu$ and positive definite matrix $\bSigma$ and let $\bx_1,\ldots,\bx_n$ be a random sample independent vectors such that $\bx_i\sim\mathcal{N}_k(\bmu,\bSigma)$ for $i=1,\ldots,n.$ and $n > k$. Then it holds that
\begin{enumerate}[i)]
\item The probability that  the estimator of portfolio based on (\ref{problem1}) is not mean-variance efficient is given by
\begin{eqnarray}\label{bm0}
\bP(\hat{\alpha}^{-1}_{1}<0)&=&\frac{n(n-k+1)}{(n-1)(k-1)}\int\limits_0^{+\infty}\left(1-\Phi\left(\lambda_{1}
\frac{1}{\sqrt{\frac{1}{n}+\frac{1}{n-1}y}}\right)\right)\nonumber\\
&\times&f_{k-1,n-k+1,ns}(\frac{n(n-k+1)}{(n-1)(k-1)}y)dy\,,
\end{eqnarray}
where $\lambda_{1}=\frac{\mu_0-R_{GMV}}{\sqrt{V_{GMV}}}=\alpha^{-1}_{1}\frac{s}{\sqrt{V_{GMV}}}.$
\item The probability that the estimator of portfolio based on (\ref{problem3}) is not mean-variance efficient is given by
\begin{eqnarray}\label{al}
\bP(\hat{\alpha}^{-1}_{3}<0)&=&\frac{n(n-k+1)}{(n-1)(k-1)}\int\limits_0^{+\infty}\left(1-\Phi\left(\lambda_{3}
\frac{1}{\sqrt{\frac{1}{n}+\frac{1}{n-1}y}}\right)\right) \nonumber\\
&\times&f_{k-1,n-k+1,ns}(\frac{n(n-k+1)}{(n-1)(k-1)}y)dy\,,
\end{eqnarray}
where $\lambda_{3}=\frac{\tilde{\alpha}^{-1}-1-R_{GMV}}{\sqrt{V_{GMV}}}=\alpha^{-1}_{3}\frac{1+s}{\sqrt{V_{GMV}}}.$
\end{enumerate}
\end{theorem}

Note that the probabilities of Theorem 2 only differ from each other by the parameters $\lambda_{1}$ and $\lambda_{3}$.

Next, we consider the probability of mean-variance inefficiency given in Theorem 2. In Figure 2 $\bP(\hat{\alpha}^{-1}_{1}<0)$ is plotted as a function of $\lambda_{1}$ for $s\in \{0.05, 0.25, 1.25\}$ with $R_{GMV}$ and $V_{GMV}$ as in (\ref{param1}). We note that the probability is a decreasing function in $\lambda_{1}$. If $\lambda_{1}=0 \,(\Leftrightarrow\alpha_{1}=\infty)$ then the probability that the estimated solutions of (\ref{problem1}) and (\ref{problem3}) are lying on the upper or the lower part of the estimated parabola is the same. For $\lambda_{1}<0\,(\Leftrightarrow\alpha_{1}<0$) the probability of getting a portfolio from the lower part of the estimated efficient frontier increases. Conversely, for positive values of $\lambda_{1}$ it decreases. If we increase $s$ then the probability becomes larger for $\lambda_{1}<0$ and smaller otherwise. Moreover, if $s$ tends to zero, what can be usually observed for real data, then the calculated probabilities tend to $0.5$ for all $\lambda_{1}$. This result is very unpleasant for investors of types making use of (\ref{problem1}) and (\ref{problem3}). In order to reduce the probability of choosing an inefficient portfolio, the investor following (\ref{problem1}) has to choose $\mu_0$ sufficiently larger than $R_{GMV}$ to ensure a large value of $\lambda_1$. Similarly, the investor using (\ref{problem3}) has to choose the coefficient $\tilde{\alpha}$ sufficiently small with respect to $R_{GMV}+1$ to make $\lambda_3$ large. But how large must be $\mu_0$ and how small must be $\tilde{\alpha}$? In this case we can construct only confidence intervals or, equivalently, provide tests for both parameters $\lambda_1^{-1}$ and $\lambda_3^{-1}$.
\begin{figure}[h!]
\begin{center}%
\includegraphics[width=0.9\textwidth]{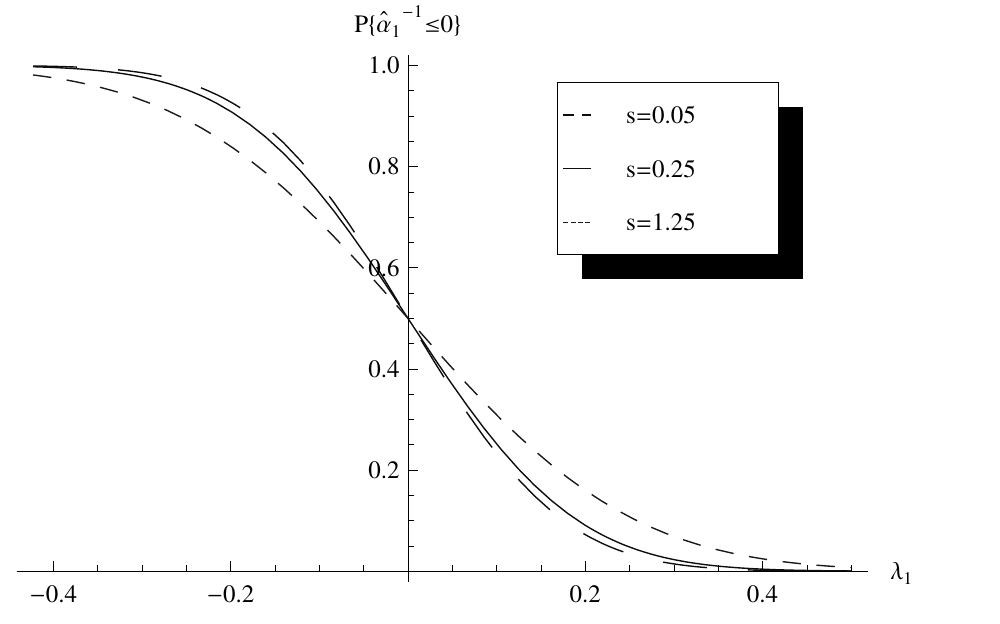}
\end{center}
\caption{{\it The probability that the estimated portfolio based on (\ref{problem1}) is not mean-variance efficient
for $s\in \{0.05, 0.25, 1.25\}$.}}%
\label{Fig:his}%
\end{figure}

\subsection{Test theory}

\noindent In this section we derive an exact one-sided test for $\alpha^{-1}_{1}$ and $\alpha^{-1}_{3}$. The aim is to provide a statistical justification of the hypothesis that the obtained solution of (\ref{problem1}) or/and (\ref{problem3}) for a given value of $\mu_0$ and $\tilde{\alpha}$, correspondingly, is mean-variance efficient, i.e. it lies on the efficient frontier. In literature several tests on mean-variance efficiency have already been proposed. Assuming that the asset returns are independently and normally distributed Gibbons et al. (1989) suggested an $F$-test for the mean-variance efficiency of a given portfolio. More recently, Britten-Jones (1999) derived a test for the tangency portfolio weights, while a test for the GMV portfolio is proposed by Bodnar and Schmid (2008a).

First, for a fixed value of $\mu_0$ we derive a test for $\alpha^{-1}_{1}$. We are interested in the testing problem
\begin{equation}\label{t12}
H_0:~~ \alpha^{-1}_{1}\leq0 ~~~~\text{against}~~~~ H_1:~~ \alpha^{-1}_{1}>0.
\end{equation}
If the null hypothesis in (\ref{t12}) is rejected we ensure that the corresponding solution lies on the efficient frontier. From the definition of $\alpha^{-1}_{1}$ the hypotheses in (\ref{t12}) are equivalent to
\begin{equation} \label{test12}
H_0:~~ R_{GMV}\geq\mu_0 ~~~~\text{against}~~~~ H_1:~~ R_{GMV}<\mu_0\,,
\end{equation}
which can be tested by applying the statistic of the test for the expected return of the $GMV$ portfolio as suggested in Bodnar and Schmid (2009). The test statistic is given by
\begin{equation}\label{stat12}
T_{1}=\sqrt{\frac{n}{\hat{V}_{GMV}}}\sqrt{\frac{n-k}{n-1}}\frac{\hat{R}_{GMV}-\mu_0}{\sqrt{1+(n/n-1)\hat{s}}}\,.
\end{equation}
Bodnar and Schmid (2009) derived the distributions of (\ref{stat12}) under both $H_0$ and $H_1$. It holds that under the null hypothesis $T_{1}\sim t_{n-k}$. Consequently, we accept the hypothesis that the solution of (\ref{problem1}) for a given value of $\mu_0$ is mean-variance efficient with the significance level $\beta$ if $T_{1}<-t_{n-k;1-\beta}$. Note that the test (\ref{test12}) is the first test where the mean-variance efficiency of the portfolio can be accepted. In previous approaches the efficiency was always specified under $H_0$. As a result it was only possible to reject the portfolio efficiency.

Under $H_1$ the exact density of $T_{1}$ is given by (Bodnar and Schmid (2009, Proposition 1))
\begin{equation}\label{denT}
f_{T_{1}}(x)=\frac{n(n-k+1)}{(k-1)(n-1)}\int\limits_0^\infty f_{t_{n-k,\delta(y)}}(x)f_{F_{k-1,n-k+1,ns}}\left(\frac{n(n-k+1)}{(k-1)(n-1)}y\right)dy\,,
\end{equation}
 with $\delta(y)=\frac{\sqrt{n}\lambda_{1}}{\sqrt{1+(n/n-1)y}}$ where $\lambda_{1}$ and $s$ are defined in Theorem 2i) and in (\ref{ef}), respectively. The symbol $f_{t_{p,\gamma}}(x)$ denotes the density of the non-central $t$-distribution with $p$ degrees of freedom and the non-centrality parameter $\gamma$. The expression (\ref{denT}) is used to study the power of the test for the testing problem given in (\ref{t12}). It must be emphasized that the power function of this test depends on $\bmu$ and $\bSigma$ only over $\lambda_{1}$ and $s$. This property dramatically simplifies the analysis of the test which can be easily performed in many mathematical packages, like e.g. Mathematica.

A similar test can be derived for $\alpha^{-1}_{3}$. In this case our aim is to test
\begin{equation}\label{t23}
H_0:~~ \alpha^{-1}_{3}\leq0 ~~~~\text{against}~~~~ H_1:~~ \alpha^{-1}_{3}>0.
\end{equation}
for a fixed value $\tilde{\alpha}$. The equivalent hypotheses are given by
\begin{equation}\label{test23}
H_0:~~ R_{GMV}\geq\tilde{\alpha}^{-1}-1 ~~~~\text{against}~~~~ H_1:~~ R_{GMV}<\tilde{\alpha}^{-1}-1.
\end{equation}
This test can be performed by applying the test statistic
\begin{equation}\label{stat23}
T_{3}=\sqrt{\frac{n}{\hat{V}_{GMV}}}\sqrt{\frac{n-k}{n-1}}\frac{\hat{R}_{GMV}-(\tilde{\alpha}^{-1}-1)}{\sqrt{1+(n/n-1)\hat{s}}}
\end{equation}
which is also $t_{n-k}$-distributed under $H_0$. The power function of this test is given by (\ref{denT}) where $\lambda_{1}$ has to be replaced by $\lambda_{3}$ as given in Theorem 2.

\begin{figure}[h!]
\begin{center}
\includegraphics[width=0.9\textwidth]{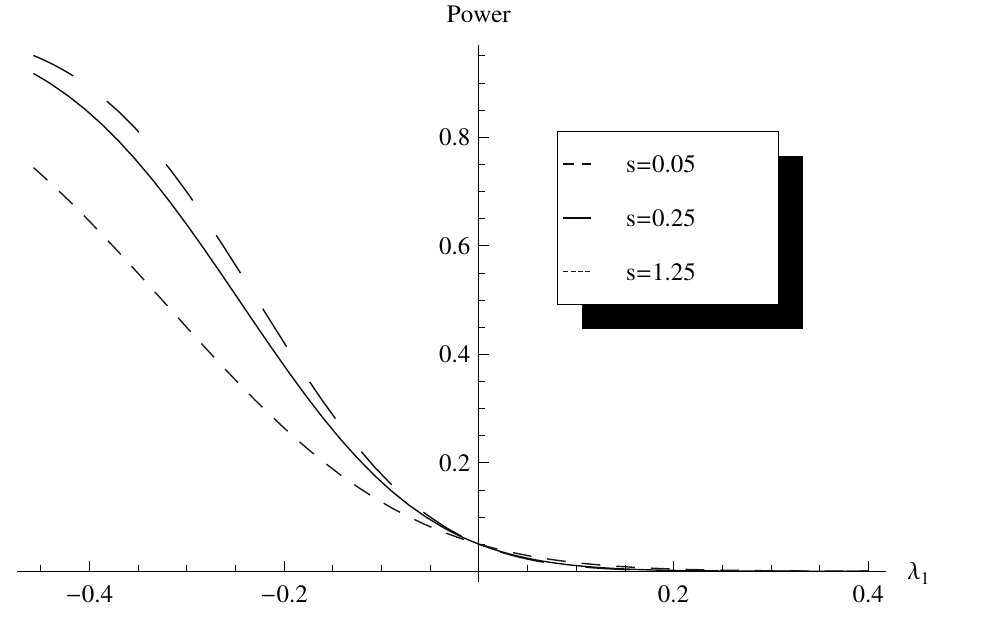}
\end{center}
\caption{Power function of the test (\ref{t12}) as a function of $\lambda_{1}$ for $s\in\{0.05,0.25,1,25\}$ with $n=60$, and $k=5$.}
\label{Fig:Pow}
\end{figure}

In Figure \ref{Fig:Pow} we present the power function of the test (\ref{t12}) as a function of $\lambda_{1}$ for $s\in\{0.05,0.25,1,25\}$. It is noted that the suggested test is powerful enough to reject the null hypothesis even for small negative values of $\lambda_{1}$. When we consider the QU optimization problem, i.e, the test (\ref{t23}), the power function has a very similar behavior as in Figure \ref{Fig:Pow}.

\section{Empirical Illustration}
In this section we apply the obtained results from Section 3 to a real data set. We consider monthly data from Morgan Stanley Capital International for the equity market returns of five developed countries (UK, Germany, USA, Canada, and Switzerland). Based on these data the parameters of the efficient frontier are chosen as
\begin{equation}\label{param2}
\hat{R}_{GMV}=0.0145664, \quad \hat{V}_{GMV}=0.0010337, \quad \text{and} \;\; \hat{s}=0.221457 \,.
\end{equation}

Our aim is to examine the probabilities of receiving a mean-variance inefficient portfolio by solving the optimization problems (\ref{problem1}) and (\ref{problem3}) according to the underlying data. These probabilities possess a very important information for the investor. Solving the equivalent optimization problem he wants to get a solution lying on the efficient frontier. However, if the probability of receiving an inefficient portfolio is high, then the two optimization problems are actually not stochastically equivalent. As a result, the application of (\ref{problem1}) and/or (\ref{problem3}) can lead to significant unexpected losses. The question is how large is the borderline for an inefficient frontier at all, i.e., how large is the probability of investing inefficiently by using (\ref{problem1}) and/or (\ref{problem3}). This question is answered in Theorem 2 of Section 3.2.

We apply the suggested testing procedures of Section 3.2 to the above mentioned empirical data set. The following step-by-step procedure has to be performed:
\begin{enumerate}
\item First, fix some values of $\bmu_0$ and $\tilde{\alpha}$ in the problems (\ref{problem1}) and (\ref{problem3}), respectively.
\item Take a sample of size $n$ from the $k$ assets and using (\ref{est}) and (\ref{estf}) calculate the sample parameters of the efficient frontier $\hat{R}_{GMV}$, $\hat{V}_{GMV}$, and $\hat{s}$ (For the considered data it holds that $k=5$, $n=60$, and the values of $\hat{R}_{GMV}$, $\hat{V}_{GMV}$, and $\hat{s}$ are given in (\ref{param2})).
\item For already found $\hat{R}_{GMV}$, $\hat{V}_{GMV}$, and $\hat{s}$ calculate the test statistics $T_1$ and $T_3$ given in (\ref{stat12}) and (\ref{stat23}), respectively.
\item For the fixed confidence level $\beta$ compare the values of $T_1$ and $T_2$ with corresponding quantile of $t$-distribution $-t_{n-k;1-\beta}$.
\item If $T_{1} (T_3)<-t_{n-k;1-\beta}$ then we accept $H_1$, i.e. the solution is mean-variance efficient, otherwise we do not reject $H_0$.
\item Go to the step 1 and repeat the procedure for different values of $\bmu_0$ and $\tilde{\alpha}$.
\end{enumerate}
The given algorithm is computationally easy to implement and can be performed in R or other statistical software.

The results of the M optimization problem are given in Figure 4a and for the QU optimization procedure in Figure 4b. In both cases we choose $\beta=0.05$.
The part of the parabola where the alternative hypothesis is accepted is given by the dashed line. In the case of (\ref{problem1}) we are able only to reject the null hypothesis for $\mu_0>0.0224823$. For the problem (\ref{problem3}) the necessary condition to ensure that its solution is mean-variance efficient is $\tilde{\alpha}^{-1}>1.0224823$ or, equivalently, $\tilde{\alpha}<0.978012$.

\begin{figure}[ptbh]
\begin{tabular}{p{8.0cm} p{8.0cm}}
$\text{a) \textit{M optimization problem}} \hspace{2.6cm}$
\includegraphics[width=7.9cm]{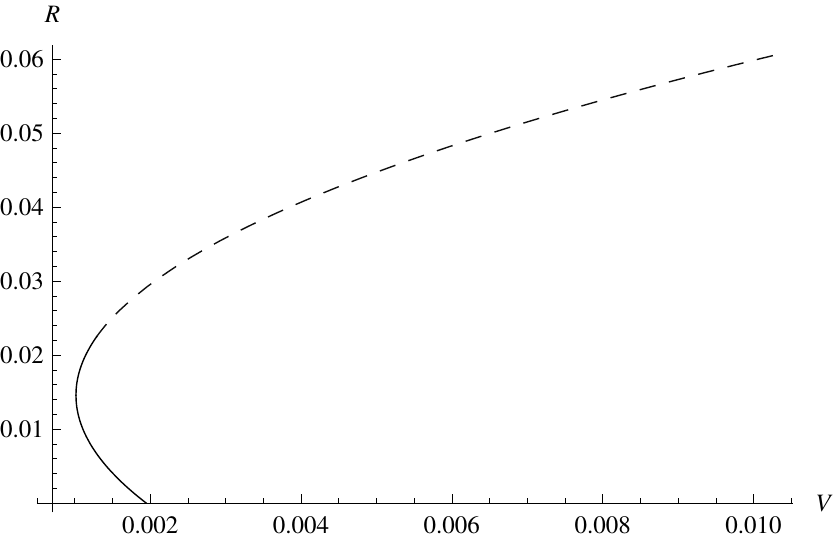}
& $\text{b) \textit{QU optimization problem}} \hspace{2.6cm}$ \includegraphics[width=7.9cm]{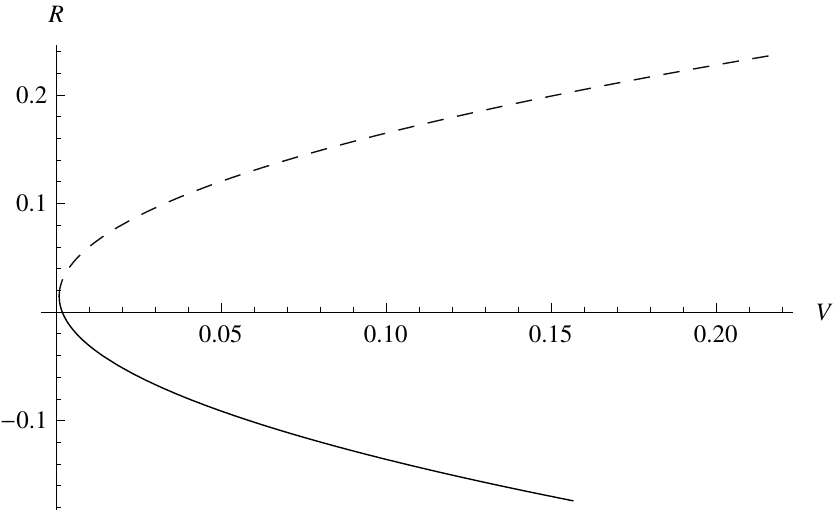}\\
\end{tabular}
\caption{Results of tests (\ref{t12}) and (\ref{t23}) are presented for different values of $\mu_0>0$(part a)) and
$\tilde{\alpha}>0$(part b)) at the significance level $\beta=0.05$. In the dash lines in both the figures the area
is shown where the alternative hypothesis of the mean-variance efficiency is accepted.}
\label{Fig:test}
\end{figure}

The above discussed testing procedures possess several nice interpretation.
They present us an intuition about how large the values of $\mu_0$ and $\tilde\alpha$ have to be chosen in order to guarantee investors using (\ref{problem1}) and (\ref{problem3}) to get an efficient portfolio and, consequently, avoid unexpected losses. Moreover, the importance of these tests seems to increase in the multi-period portfolio context where the above mentioned optimization problems are commonly used and the unexpected loss can be accumulated from one period to another. Thus, for investors using (\ref{problem1}) and (\ref{problem3}) it is reasonable to correct the parameters $\mu_0$ and $\tilde{\alpha}$ before constructing the portfolio. No correction is necessary for investor using the optimization problem (\ref{problem2}).

The assumption of normality and independence of the asset returns is, of course, very restrictive. However, assuming more complicated distributions or processes it is very difficult to get such expressions as derived in Section 3. The real asset returns are heavily tailed with a high peak in the center so that the probability of getting an inefficient portfolio can be even larger. That is why we consider the above mentioned procedure as a benchmark how the probability of mean-variance inefficiency can be determined.

\section{Summary}
\noindent In 1952 G. Markowitz suggested a novel approach how a optimal portfolio can be constructed. The idea of Markowitz was to minimize the portfolio variance for a given level of return. This optimization problem is known in literature as the mean-variance analysis. Recently, other optimization procedures have been suggested (see, e.g., Ingersoll (1987), Brandt and Santa-Clara (2006), Bodnar and Schmid (2008b, 2009)) which are based on quadratic utility functions.

In the present paper we compare three approaches with each other. It turns out that under some conditions they are mathematically equivalent but not always mean-variance efficient. It holds that only all optimal portfolios obtained by maximizing the mean-variance utility function are mean-variance efficient. Conditions are derived under which the solutions of the other considered optimizations problems lay on the efficient frontier. These conditions, however, cannot be checked in a practical situation because they all depend on the unknown parameters of the asset returns. We provide a comprehensive treatment of this problem by deriving the exact expressions of the probabilities that the estimated solutions of the Markowitz problem and the one obtained by maximizing the quadratic utility are mean-variance efficient. Because the derived probabilities deviate from one we conclude that the three optimization problems are not stochastically equivalent. Finally, exact tests for the mean-variance efficiency of the obtained solutions are developed.

\section{Appendix}
\noindent In this section the proof of Theorem 1 is given.

\noindent \textbf{Proof of Theorem 1:}

\begin{enumerate}[i)]

\item From Merton (1972) the solution of (\ref{problem1}) is given by
\begin{eqnarray*}
\bw=\frac{a-\mu_0b}{ac-b^2}\bSigma^{-1}\bii+\frac{(\mu_0c-b)}{ac-b^2}\bSigma^{-1}\bmu\,,
\end{eqnarray*}
where the constants $a$, $b$, $c$, $R_{GMV}$, $V_{GMV}$, and $s$ are defined at the beginning of Section 2. With some mathematical calculations, we get
\begin{eqnarray*}
\bw&=&\frac{\bSigma^{-1}\bii}{\bii^\prime\bSigma^{-1}\bii}+\left(\frac{a-\mu_0b}{a-b^2/c}-1\right)\frac{\bSigma^{-1}\bii}{\bii^\prime\bSigma^{-1}\bii}+\frac{(\mu_0c-b)}{ac-b^2}\bSigma^{-1}\bmu\\
&=&\frac{\bSigma^{-1}\bii}{\bii^\prime\bSigma^{-1}\bii}+\left(\frac{b^2/c-\mu_0b}{a-b^2/c}\right)\frac{\bSigma^{-1}\bii}{\bii^\prime\bSigma^{-1}\bii}+\frac{(\mu_0c-b)}{ac-b^2}\bSigma^{-1}\bmu\\
&=&\frac{\bSigma^{-1}\bii}{\bii^\prime\bSigma^{-1}\bii}-\frac{\mu_0-R_{GMV}}{s}R_{GMV}\bSigma^{-1}\bii+\frac{\mu_0-R_{GMV}}{s}\bSigma^{-1}\bmu\\
&=&\frac{\bSigma^{-1}\bii}{\bii^\prime\bSigma^{-1}\bii}+\frac{\mu_0-R_{GMV}}{s}\bQ\bmu \,.
\end{eqnarray*}

The last expression is the solution of (\ref{problem2}) with $\alpha^{-1}={(\mu_0-R_{GMV})}/{s}$ (see, e.g., Bodnar and Schmid (2008b, p. 1003)). The statement of the theorem is proved.

\item The solution of (\ref{problem3}) is given by
\begin{eqnarray}\label{wA}
\bw&=&\displaystyle\frac{\bA^{-1}\bii}{\bii^\prime\bA^{-1}\bii}+\tilde{\alpha}^{-1}\left(\bA^{-1}-\displaystyle\frac{\bA^{-1}\bii\bii^\prime\bA^{-1}}{\bii^\prime\bA^{-1}\bii}
\right)\tbm\,.
\end{eqnarray}

The application of the Sherman-Morrison formula (Harville (1997, Theorem 18.2.8)) leads to
\begin{eqnarray*}
\bw&=&\frac{\bSigma^{-1}\bii(1+\tbm^\prime\bSigma^{-1}\tbm)-\bSigma^{-1}\tbm\tbm^\prime\bSigma^{-1}\bii}
{\bii^\prime\bSigma^{-1}\bii(1+\tbm^\prime\bSigma^{-1}\tbm)-(\bii^\prime\bSigma^{-1}\tbm)^2}
+\tilde{\alpha}^{-1}\left(\bSigma^{-1}\tbm-\frac{\bSigma^{-1}\tbm\tbm^\prime\bSigma^{-1}\tbm}{1+\tbm^\prime\bSigma^{-1}\tbm}\right.\\
&-&\left.\frac{\bii^\prime \bSigma^{-1}\tbm(1+\tbm^\prime\bSigma^{-1}\tbm)-\bii^\prime\bSigma^{-1}\tbm\tbm^\prime\bSigma^{-1}\tbm}
{\bii^\prime\bSigma^{-1}\bii(1+\tbm^\prime\bSigma^{-1}\tbm)-(\bii^\prime\bSigma^{-1}\tbm)^2}
\left(\bSigma^{-1}\bii-\frac{\bSigma^{-1}\tbm\tbm^\prime\bSigma^{-1}\bii}{1+\tbm^\prime\bSigma^{-1}\tbm}\right)\right)\\
&=&\frac{\bSigma^{-1}\bii}{\bii^\prime\bSigma^{-1}\bii}\frac{1+\tbm^\prime\bSigma^{-1}\tbm}{1+\tbm^\prime\bQ\tbm}
-\frac{\tbm^\prime\bSigma^{-1}\bii}{\bii^\prime\bSigma^{-1}\bii(1+\tbm^\prime\bQ\tbm)}\bSigma^{-1}\tbm
+\tilde{\alpha}^{-1}\left(\frac{\bSigma^{-1}\tbm}{1+\tbm^\prime\bSigma^{-1}\tbm}\right.\\
&-&\left.\frac{\bii^\prime \bSigma^{-1}\tbm}
{\bii^\prime\bSigma^{-1}\bii(1+\tbm^\prime\bQ\tbm)}\left(\bSigma^{-1}\bii
-\frac{\bSigma^{-1}\tbm\tbm^\prime\bSigma^{-1}\bii}{1+\tbm^\prime\bSigma^{-1}\tbm}\right)\right)
\end{eqnarray*}

Because ${\bii^\prime \bSigma^{-1}\tbm}/{\bii^\prime \bSigma^{-1}\bii}=1+R_{GMV}$ and $\tbm^\prime\bQ\tbm=s$ we get
\begin{eqnarray*}
\bw&=&\frac{\bSigma^{-1}\bii}{\bii^\prime\bSigma^{-1}\bii}-\frac{\tilde{\alpha}^{-1}-1-R_{GMV}}{1+s}(1+R_{GMV})\bSigma^{-1}\bii
+\frac{\tilde{\alpha}^{-1}-1-R_{GMV}}{1+s}\bSigma^{-1}\tbm\\
&=&\frac{\bSigma^{-1}\bii}{\bii^\prime\bSigma^{-1}\bii}+\frac{\tilde{\alpha}^{-1}-1-R_{GMV}}{1+s}\bQ\tbm
=\frac{\bSigma^{-1}\bii}{\bii^\prime\bSigma^{-1}\bii}+\frac{\tilde{\alpha}^{-1}-1-R_{GMV}}{1+s}\bQ\bmu\,,
\end{eqnarray*}
where the last equality follows from $\bQ\tbm=\bQ\bmu$. The theorem is proved.
\end{enumerate}

\section*{Acknowledgements}

\noindent The authors are thankful to the Referees and the
Editor for their suggestions which have improved the presentation in
the paper.

\end{document}